# $x$-index: A Fantastic New Indicator for Quantifying a Scientist's Scientific Impact

Xiaojun Wan

Institute of Computer Science and Technology, The MOE Key Laboratory of Computational Linguistics, Peking University, Beijing, China

wanxiaojun@pku.edu.cn

**Abstract**

$h$-index has become the most popular indicator for quantifying a scientist's scientific impact in various scientific fields. $h$-index is defined as the largest number of papers with citation number $\geq h$ and it treats each citation equally. However, different citations usually come from different papers with different influence and quality, and a citation from a highly influential paper is a greater recognition of the target paper than a citation from an ordinary paper. Based on this assumption, we proposed a new indicator named $x$-index to quantify a scientist's scientific impact by considering only the citations coming from influential papers. $x$-index is defined as the largest number of papers with influential citation number $\geq x$, where each influential citation comes from a paper for which the average ACNPP (Average Citation Number Per Paper) of its authors $\geq x$. Through analysis on the APS dataset, we found that the proposed $x$-index has better ability to discriminate between Physics Prize Winners and ordinary physicists.

## Introduction

How to reasonably quantify the scientific impact of a scientist is an important and challenging problem in the fields of information sciences and scientometrics. The pursuit of high scientific impact is the long-term goal of a scientist in his or her research career, and the scientific impact of a scientist have great impact on his or her career activities, such as funding application, position promotion and award winning. Since research papers or articles published by a scientist are the most obvious research outputs and achievements of the scientist, the scientific impact of a scientist is usually measured by analyzing the impact of his or her published research papers or articles.

Till now, many different indicators have been developed to quantify the scientific impact of different scientists based on their published papers in the scientific community. Among the various indicators, the most popular one is $h$-index (Hirsch 2005). $h$-index is defined as the largest number of papers with citation number $\geq h$. It relies on the number of the citations received by each article and if a scientist has published more papers with more citations received, then the scientist's $h$-index is high. We can see $h$-index treats each received citation equally, however, different citations usually come from different citing papers, some of which are more influential than others. A citation from a highly influential citing paper is called an influential citation, which is usually a greater recognition of the cited paper than a citation from an ordinary citing paper. Therefore, we aim to consider only the influential citations received by a scientist to evaluate the scientist's scientific impact. We use the average ACNPP (Average Citation Number Per Paper) of authors for a citing paper to assess the paper's influence. Based on this assumption, we propose a new indicator named $x$-index to quantify a scientist's scientific impact. $x$-index is defined as the largest number of papers with influential citation number $\geq x$, where each influential citation comes from a citing paper for which the average ACNPP (Average Citation Number Per Paper) of its authors $\geq x$. Our proposed $x$-index is a variant of $h$-index with citation quality control and $x$-index is more stringent than $h$-index.

Through quantitative analysis on the APS dataset, we found that the proposed *x*-index has better ability to discriminate between Physics Prize Winners and ordinary physicists than h-index and several other indicators. .

## Related Work

Up to now, a few various indicators have been proposed for evaluating the scientific impact of a scientist or researcher based on article-to-article citations. The most popular one is *h*-index, which is defined as the number of papers with citation number $\geq h$, which has been the most widely used indicator to characterize the scientific output of a researcher (Hirsch 2005). The index has already been used in practical services such as Google Scholar and ISI Web of Science. According to a stochastic model proposed in (Burrell 2007), the *h*-index is approximately linear in career length, log publication rate and log citation rate. The predictive power of the *h* index has been empirically compared with other indicators (total citation count, citations per paper, and total paper count), and it has been shown to be better than other indicators in predicting future scientific achievement (Hirsch 2007).

In addition, a few variant indicators have been proposed on the basis of *h*-index, including *g*-index (Egghe 2006), *a*-index (Jin 2006), *hg*-index (Alonso et al. 2009), *r*-index (Jin et al. 2007), *m*-index (Bornmann et al. 2008), $h_a$-index (Van Eck & Waltman 2008), $h^{(2)}$-index (Kosmulski 2006), $h^n$-index (Sidiropoulos et al. 2007), $h_{rat}$-index (Ruane & Tol 2008), *ar*-index (Jin 2007), $h_w$-index (Egghe and Rousseau 2008), *p*-index (Prathap & Gupta, 2009), *j*-index (Todeschini 2011), and so on. Almost all these indicators make some modifications on the *h*-index. For example, the *g*-index gives more weight to highly cited papers and it is defined as the highest number *g* of papers that together received $g^2$ or more citations. A few studies have been conducted to review or compare some of these indicators (Alonso et al. 2009; Costas and Bordons 2007; Bornmann, Mutz and Daniel 2008; Bar-Ilan 2008). However, all the above indicators do not consider the citation quality and the citations coming from different papers written by different researchers are treated uniformly.

**Our Proposed Indicator**

Since our proposed indicator is a Hirsch-style indicator, we have to take a close look at $h$-index. The original definition of the $h$-index is as follows (Hirsch 2005):

"A scientist has index $h$ if $h$ of his or her $N_p$ papers have at least $h$ citations each and the other ($N_p - h$) papers have $\leq h$ citations each."

$h$-index is computed based on the citations received by each paper published by a scientist and the intuition behind $h$-index is that if a scientist has more influential papers, the scientist has high $h$-index. The influence of a paper is measured by the number of citations it receives.

We can see that different citations received by a paper is treated equally in the computation of $h$-index. However, different citations in a paper usually come from different citing papers, and these papers usually have different influences. For example, a citation coming from a citing paper written by a senior scientist is usually considered a greater recognition of the cited paper than a citation coming from a citing paper written by a junior student. That's to say, different citations in a paper are not equal recognition of the paper and the different recognition can be leveraged to quantify a scientist's scientific impact.

In order to reflect the different influences of different citations received by a scientist, we propose a new indicator – $x$-index to quantify a scientist' scientific impact.

***x*-index**: A scientist has index $x$ if $x$ of his or her $N_p$ papers have received at least $x$ influential citations each and the other ($N_p - x$) papers have received $\leq x$ influential citations each, where an influential citation comes from a citing paper whose authors have an average ACNPP (Average Citation Number Per Paper) $\geq x$.

Our proposed *x*-index considers only the influential citations received by a scientist, and ignores all other citations. The influence of a citation is measured by the scientific impact of the citing paper proposing the citation, and the scientific impact of a paper is measured by the average ACNPP (Average Citation Number Per Paper) of its authors. Note that ACNPP is a simple and widely-used metric to evaluate the scientific impact of an author, and it is simply computed by dividing the total citation number received by the author by the total paper number of the author.

The intuition behind *x*-index is that if a scientist has more papers cited by more influential scientists, the scientist has high *x*-index. Seen from the definition, *x*-index is a more stringent indicator than *h*-index and the *x*-index of a scientist is always less than or equal to the *h*-index of the scientist. In comparison with *h*-index, *x*-index focuses more on the citation quality than the citation number, and thus it is deemed to have better ability to quantify a scientist's scientific impact.

## Quantitative Analysis

In the experiments, we use the APS dataset provided by American Physical Society. The dataset is publicly available[1] and it contains over 450,000 articles published from 1893 through 2009 in a variety of premier physical journals including Physical Review Letters, Physical Review and Reviews of Modern Physics. The dataset contains the basic metadata of all articles, including the following fields: DOI, title, authors, affiliations, journal, volume, issue, publication history, and etc. The dataset also contains all pairs of APS articles that cite each other. For instance, if article A cites article B, there will be an entry in the dataset consisting of the pair of DOIs for A and B.

---

[1] https://publish.aps.org/datasets

In order to compare our proposed *x*-index and other indicators, we have to build a benchmark. Similar to (Jiang et al. 2013), we evaluate the indicators by comparing their ability to discriminate Physics Prize Winners and ordinary physicists. The prize winners have been widely recognized in the physical community and they usually have higher scientific impact than ordinary physicists. We rank the physicists according to each indicator, and the more prize winners are ranked higher than the ordinary physicists by an indicator, the better the indicator is. We collect winners for Physics Nobel Prize and seven major physics prizes (Boltzmann Medal, Dannie Heineman Prize, Matteucci Medal, Max Planck Medal, Sakurai Prize, Wolf Prize and Enrico Fermi Award) after 1960. These prize winners are usually of high scientific impact than ordinary physicists and they are expected to be ranked higher than ordinary physicists.

In the dataset, sometimes a name may refer to different persons and sometimes different names may refer to the same person due to name ambiguities. As in (Jiang et al. 2013), we do not do person name disambiguation because this task requires more information which is not available in the APS dataset and the performance of existing name disambiguation algorithms is far from satisfactory. Instead, we simply abbreviate each author's full name into a canonical form consisting of his/her last name and the first character of his/her first name. For example, "John C. Mather" will be abbreviated into "J. Mather". In this way, different names referring to the same person will be treated properly. Sometimes different persons sharing the same name may be improperly treated as one person, but fortunately, such situations do not occur frequently, especially for the prize winners. After abbreviation, there are totally 183459 authors in the APS dataset, and 162736 out of them have been cited at least once. We have collected 122 Physics Nobel Prize Winners and 315 other major physics prize winners, and there is a total of 403 prize winners since some Nobel Prize winners also won other major physics prizes.

All the physicists in the APS dataset are ranked according to each indicator, and if there is a tie between multiple physicists, we re-rank them by the total citation numbers of them. We compare *x*-index and other indicators in an information retrieval environment. We considered prize winners as the gold-standard set, ranked all the physicists according to an indicator, and then used typical information retrieval evaluation metrics to evaluate the ranked list against the gold-standard set. We used the following metrics for evaluation:

**P@N**: It measures the precision of the top $N$ physicists in the ranked list ($N$ is typically 10~100), which is computed as follows:

$$P@N = \frac{\text{The number of prize winners in the top } N \text{ physicists}}{N}$$

**AP**: It first determines precision at each position when each prize winner is found in the ranked results, and if a prize winner in the gold-standard set is not found, the corresponding precision is 0, and finally the precision values are averaged.

In addition to $h$-index, we also compare our proposed $x$-index with ACNPP (Average Citation Number Per Paper), TCN (Total Citation Number), and TPN (Total Paper Number).

The comparison results based on different metrics are shown in Tables 1 and 2. In Table 1, we only consider Nobel Prize winners as the gold-standard set, while in Table 2, both Nobel Prize winners and other prize winners are considered the gold-standard set.

Seen from Tables 1 and 2, both $x$-index and $h$-index outperform the simple indicators, including ACNPP, TCN and TPN. More interestingly, the ranked results based on $x$-index perform much better than that based on $h$-index over all evaluation metrics, and the precision scores have even been improved by 50%~100%. The results show that $x$-index can be used to put more prize winners in the top of the ranked list. Considering the fact that the prize winners usually have higher scientific impact than ordinary physicists, $x$-index has shown the superior ability to discriminate prize winners from ordinary physicists. Therefore, $x$-index has sensible advantages over $h$-index.

Table 3 gives the top 20 physicists ranked by *x*-index and *h*-index in the APS dataset, respectively. The *x*-index and *h*-index scores of the physicists are also presented in the table. In the top 20 physicists ranked by *x*-index, four out of them are Novel Prize winners (including P. Anderson, S. Weinberg, W. Ketterle, and F. Wilczek), and five out of them are winners of other major physics prizes (including M. Fisher, S. Weinberg, B. Halperin, P. Zoller, and F. Wilczek), and there is a total of seven prize winners. Whereas, in the top 20 physicists ranked by *h*-index, only two out of them are Novel Prize winners (including P. Anderson, and S. Weinberg), and four out of them are winners of other major physics prizes (including M. Fisher, S. Weinberg, B. Halperin, and P. Zoller), and there is a total of five prize winners. The results demonstrate that *x*-index can really rank more prize winners ahead of ordinary physicists.

We can also see from Table 3 that the *x*-index scores are generally much lower than the *h*-index scores, which can be attributed to the fact that *x*-index is a more stringent indicator than *h*-index by adding the constraints of citation quality. For example, P. Anderson's *h*-index is 53, while his *x*-index is dropped to 23. S. Kim has the highest *h*-index while he has been squeezed out from the top list ranked by *x*-index. Overall, our proposed *x*-index has better ability to indicate the scientific impact of a scientist than *h*-index.

Table 1

*Comparison results based on Nobel Prize Winners*

|          | *x*-index | *h*-index | ACNPP  | TCN    | TPN    |
|----------|-----------|-----------|--------|--------|--------|
| **P@10** | **0.3**   | 0.2       | 0      | 0      | 0      |
| **P@20** | **0.2**   | 0.1       | 0      | 0.05   | 0      |
| **P@30** | **0.2**   | 0.0667    | 0      | 0.1    | 0.0333 |
| **P@40** | **0.175** | 0.05      | 0      | 0.075  | 0.025  |
| **P@50** | **0.2**   | 0.08      | 0      | 0.06   | 0.02   |
| **P@100**| **0.16**  | 0.06      | 0      | 0.04   | 0.02   |
| **AP**   | **0.0484**| 0.0206    | 0.0046 | 0.0146 | 0.0046 |

Table 2

*Comparison results based on All Prize Winners (Nobel Prize Winners and seven major prize winners)*

|  | *x*-index | *h*-index | ACNPP | TCN | TPN |
|---|---|---|---|---|---|
| **P@10** | **0.6** | 0.3 | 0 | 0 | 0 |
| **P@20** | **0.35** | 0.25 | 0.05 | 0.1 | 0 |
| **P@30** | **0.3667** | 0.2333 | 0.0333 | 0.1333 | 0.0333 |
| **P@40** | **0.3** | 0.175 | 0.025 | 0.15 | 0.025 |
| **P@50** | **0.3** | 0.2 | 0.02 | 0.18 | 0.02 |
| **P@100** | **0.25** | 0.12 | 0.02 | 0.11 | 0.02 |
| **AP** | **0.0482** | 0.0283 | 0.0119 | 0.0255 | 0.0099 |

Table 3

*Top 20 authors in the APS dataset*

*([*] indicates Nobel Prize Winner and [#] indicates Other Major Physics Prize Winner.)*

| **Ranked by *x*-index** | | **Ranked by *h*-index** | |
|---|---|---|---|
| name | *x*-index | name | *h*-index |
| *P. Anderson[*]* | 23 | S. Kim | 53 |
| V. Kostelecky | 23 | *P. Anderson[*]* | 53 |
| *M. Fisher[#]* | 22 | M. Cohen | 52 |
| *S. Weinberg[*#]* | 22 | *M. Fisher[#]* | 52 |
| P. Lee | 21 | *S. Weinberg[*#]* | 52 |
| *B. Halperin[#]* | 21 | H. Kim | 51 |
| M. Cohen | 20 | C. Wang | 49 |
| J. Cirac | 20 | P. Lee | 48 |
| *P. Zoller[#]* | 20 | L. Pfeiffer | 48 |
| *W. Ketterle[*]* | 20 | K. West | 48 |
| D. Fisher | 20 | D. Kim | 47 |
| M. Kastner | 20 | *B. Halperin[#]* | 47 |
| A. Suzuki | 19 | R. Wagner | 46 |
| L. Pfeiffer | 19 | D. Brown | 46 |
| K. West | 19 | J. Cirac | 46 |
| J. Schrieffer | 19 | *P. Zoller[#]* | 45 |
| T. Rice | 19 | S. Lee | 44 |
| D. Scalapino | 19 | J. Wang | 44 |
| *F. Wilczek*[*#] | 19 | A. Zunger | 44 |
| X. Wen | 19 | S. Louie | 43 |

## Conclusions and Future Work

In this article, we incorporated citation quality into $h$-index and proposed the new $x$-index for quantifying the scientific impact of a scientist. Based on quantitative analysis on the APS dataset, we found $x$-index has better ability to indicate the scientific impact of a physicist than $h$-index.

In future work, we will apply our proposed $x$-index to quantify the scientific impact of scientists in other disciplines to demonstrate its robustness. Furthermore, $x$-index can be easily adapted to quantify the scientific impact of journals and organizations.

## Acknowledgements

The work was supported by National Natural Science Foundation of China (NSFC) (No. 61170166 & No. 61331011) and the National High Technology Research and Development Program of China (863 Program) (No. 2012AA011101).